\documentclass[twocolumn,prd]{revtex4}
\usepackage{amsmath}
\usepackage{mathrsfs,bm,multirow}
\usepackage{longtable,lscape}
\usepackage{txfonts}
\usepackage{amssymb}
\usepackage{indentfirst}
\usepackage{graphicx,,booktabs}
\usepackage{color}
\usepackage{amssymb}
\begin{document}

\title{Reply to `` Comment on `Finding the $0^{--}$ Glueball' " and comment on `Is the exotic $0^{--}$ glueball a pure gluon state?'}
\author{Cong-Feng Qiao$^{1,3}$}
\author{Liang Tang$^{2,3}$}
\affiliation{$^1$ School of Physics, University of Chinese Academy of Sciences - YuQuan Road 19A, Beijing 100049, China\\
$^2$ Department of Physics, Hebei Normal University, Shijiazhuang 050024, China\\
$^3$ CAS Center for Excellence in Particle Physics, Beijing 100049, China}

\maketitle

In the preceding comment \cite{Pimikov:2017xap}, Pimikov {\it et al.} argued that the two-point correlation function, the correlator, on the left hand side of the sum rule tends to be negative, and then leads to an unphysical negative decay constant, with our choice of the gluonic current in \cite{Qiao:2014vva}. In fact, the minus sign can be trivially removed by giving a proper phase to the current. That is, the modified currents have the following expressions:
\begin{eqnarray}
j^A_{0^{--}}(x) & \!\!\!\! = \!\!\!\! & i g_s^3 d^{a b c} [g^t_{\alpha \beta} \tilde{G}^a_{\mu \nu}(x)][\partial_\alpha \partial_\beta G^b_{\nu \rho}(x)][G^c_{\rho \mu}(x)]\, , \label{current-A} \\
j^B_{0^{--}}(x) & \!\!\!\! = \!\!\!\! & i g_s^3 d^{a b c} [g^t_{\alpha \beta} G^a_{\mu \nu}(x)][\partial_\alpha \partial_\beta \tilde{G}^b_{\nu \rho}(x)][G^c_{\rho \mu}(x)]\, , \label{current-B} \\
j^C_{0^{--}}(x) & \!\!\!\! = \!\!\!\! & i g_s^3 d^{a b c} [g^t_{\alpha \beta} G^a_{\mu \nu}(x)][\partial_\alpha \partial_\beta G^b_{\nu \rho}(x)][\tilde{G}^c_{\rho \mu}(x)]\, , \label{current-C} \\
j^D_{0^{--}}(x) & \!\!\!\! = \!\!\!\! & i g_s^3 d^{a b c} [g^t_{\alpha \beta} \tilde{G}^a_{\mu \nu}(x)][\partial_\alpha \partial_\beta \tilde{G}^b_{\nu \rho}(x)][\tilde{G}^c_{\rho \mu}(x)]\, .\label{current-D}
\end{eqnarray}
In this case, the Wilson coefficient $a_0^i$ will reverse the sign and lead to a positive decay constant $f_G$. In all, the above currents may couple to gluonic bound states and the pure gluonic $0^{--}$ glueball masses will be deduced \cite{Qiao:2014vva}.

Following, we would like to comment on a recent paper `Is the exotic $0^{--}$ glueball a pure gluon state?' \cite{Pimikov:2016pag}, in which authors constructed a new glueball current to study the exotic $0^{--}$ glueball. The current takes following form:
\begin{eqnarray}
  J^{LO}(x) = g_s^3 d^{abc} \tilde{G}^a_{\mu \nu ; \tau_1 \tau_2 \tau_3}(x) G^b_{\nu \rho ; \tau_1 \tau_2}(x) G^c_{\rho \mu ; \tau_3}(x) . \label{eq1}
\end{eqnarray}

It is obvious that while the place of $\tilde{G}$ is changed or all $G$s are replaced by $\tilde{G}$s in current (\ref{eq1}), different  currents will be obtained, which can not be simply ignored in order to obtain a reliable conclusion.

Moreover, in the technique of QCD Sum Rules \cite{Shifman, Reinders:1984sr, P.Col, Matheus:2006xi}, it is well-known that one of the two criteria to constraint the Borel parameter $M_B^2$ is that the pole contribution should exceed that from the higher excited and continuum states. Therefore, one needs to evaluate the relative pole contribution over the total, the pole plus the higher excited and continuum($s_0 \to \infty$). In order to properly eliminate the contribution from higher excited and continuum states \cite{Reinders:1984sr, P.Col, Matheus:2006xi}, the pole contribution is generally required to be more than 50\%.

Nevertheless, in Ref. \cite{Pimikov:2016pag}, this criterion is taken as:
\begin{eqnarray}
  \frac{{\cal R}^{(SR)}_k(M^2, s_0)}{{\cal R}^{(SR)}_k(M^2, \infty)} > \frac{1}{10} \;.
\end{eqnarray}
This restriction is too weak to obtain a reliable fiducial interval for the Borel parameter, and hence makes the corresponding prediction dubious.

%%%%%%%%%%%%%%%%%%%%%%%%%%%%%%%%%%%%%%%%%%%%%%%%%%%%%%%%%%%%%%%%%%%%%%%%%%%%

%%%%%%%%%%%%%%%%%%%%%%%%%%%%%%%%%%%%%%%%%%%%%%%%%%%%%%%%%%%%%%%%%%%%%%%%%%%%

\end{document}